\documentclass[11pt]{article}

\usepackage[margin=1in]{geometry}
\usepackage[utf8]{inputenc}
\usepackage[T1]{fontenc}
\usepackage{newtxtext,newtxmath}
\usepackage{graphicx}
\usepackage{booktabs}
\usepackage{longtable}
\usepackage{array}
\usepackage{caption}
\usepackage{amsmath}
\usepackage{xcolor}
\usepackage[numbers,round]{natbib}
\usepackage{titlesec}
\usepackage{xurl}
\usepackage[colorlinks=true,linkcolor=blue,citecolor=blue,urlcolor=blue]{hyperref}

\usepackage{newunicodechar}
\newunicodechar{≤}{\ensuremath{\leq}}
\newunicodechar{≥}{\ensuremath{\geq}}
\newunicodechar{±}{\ensuremath{\pm}}
\newunicodechar{×}{\ensuremath{\times}}
\newunicodechar{®}{\textsuperscript{\textregistered}}

\usepackage{calc}
\providecommand{\real}[1]{#1}

\titleformat{\section}{\normalfont\large\bfseries}{\thesection}{1em}{}
\titleformat{\subsection}{\normalfont\normalsize\bfseries}{\thesubsection}{1em}{}

\title{\textbf{Experts Disagree on How to Fight AI Disinformation,\\
but Agree That Health and Politics Need Different Solutions}\\[4pt]
\large\normalfont Author-Formatted Version%
\thanks{This is an author-formatted version, identical in content to the
article published in the
\emph{Harvard Kennedy School Misinformation Review}, 7(4), 2026,
under a CC BY 4.0 license. The publisher's version of record is available at
DOI: \href{https://doi.org/10.37016/mr-2020-205}{10.37016/mr-2020-205}.
Replication materials: Harvard Dataverse,
\href{https://doi.org/10.7910/DVN/BXO2QA}{10.7910/DVN/BXO2QA}.}}

\author{%
Alexander Loth\thanks{Corresponding author: \href{mailto:alexander.loth@stud.fra-uas.de}{alexander.loth@stud.fra-uas.de}}\;$^{1}$ \and
Martin Kappes\;$^{1}$ \and
Marc-Oliver Pahl\;$^{2}$ \\[6pt]
\normalsize $^{1}$Faculty of Computer Science and Engineering,\\
\normalsize Frankfurt University of Applied Sciences, Frankfurt am Main, Germany\\
\normalsize $^{2}$Chair of Cybersecurity in Critical Networked Infrastructures (Cyber CNI),\\
\normalsize IMT Atlantique, Rennes, France
}

\date{July 29, 2026}

\begin{document}
\maketitle

\section{Research questions }\label{research-questions}

\begin{itemize}
\item
  How do experts perceive AI-generated disinformation threats across four modalities (text, images, audio, video) and four domains (political, health, financial, social)?
\item
  Which modality-domain combinations do experts view as most dangerous, and do threat profiles differ systematically across domains?
\item
  How do experts evaluate the effectiveness of five mitigation strategies and which do they prioritize?
\item
  What do experts identify as the most urgent near-term risks from generative AI?
\end{itemize}

\section{Research note summary}\label{research-note-summary}

\begin{itemize}
\item
  We surveyed 54 experts on AI-driven disinformation (online, July 2025--March 2026) reached via targeted recruitment with snowball extension; 454 invited, 11.9\% response rate.
\item
  Video deepfakes received the highest average threat ratings overall (averaging 6.19 on a 7-point scale), but threat patterns varied by domain: in health, AI-generated text had the highest (\emph{M} = 5.80) and video had the lowest (\emph{M} = 5.13); in finance, audio deepfakes had the highest threat ratings (\emph{M} = 5.65). Views of mitigation strategies were contested but not polarized: government regulation drew both the most ``most effective'' (30\%) and, notably, the most ``least effective'' (15\%) votes; media literacy was split 26\% ``most effective'' to 24\% ``least effective''. Effectiveness ratings were right-skewed and unimodal, suggesting disagreement is about priority and not about whether strategies work. Election interference via deepfake video was rated as the top urgent risk (78\%).
\item
  These preliminary findings suggest that respondents perceived domain-specific policy approaches as more appropriate than uniform ones. Experts consistently highlighted voice-cloning fraud as an area warranting particular regulatory attention. Because no single intervention commanded consensus, the findings suggest that layered mitigation approaches may be preferable across provenance, literacy, regulation, and platform enforcement, weighted to each domain's threat profile (see Appendix Table C4).
\end{itemize}

\section{Implications}\label{implications}

\subsection{Domain-specific threat patterns suggest tailored interventions}\label{domain-specific-threat-patterns-suggest-tailored-interventions}

Our respondents perceived distinct threat profiles in each domain: text dominated in health, audio in finance, and video in politics. Current governance instruments are not organized along those lines. The EU AI Act \citep{europarl2024aiact} and the Digital Services Act impose horizontal obligations - transparency labeling, risk assessment, systemic-risk audits - that apply identically whether the content is a fabricated medical claim or a cloned voice in a payment fraud. In the United States, oversight is fragmented across sectoral agencies, with the FDA covering health claims and the FTC covering deceptive commercial practices, and no federal AI statute in place \citep{bommasani2024fmti}. Neither the horizontal EU approach nor the sectoral U.S. patchwork tracks the domain-by-modality threat structure our respondents described.

In the political domain, where video deepfakes received the highest average threat ratings, respondents frequently identified detection and provenance standards as priority areas for future investment. As Giovanni Spitale (University of Zurich) stated, ``My biggest concern is electoral interference and more in general interference in democratic processes, because once you break that you can break each and every other element of democratic societies.'' The Coalition for Content Provenance and Authenticity (C2PA) standard responds to exactly this concern: it attaches cryptographically signed capture and edit history to a media file, so a viewer can check whether a video originated from a real camera or from a generative model. Its effectiveness depends on broad adoption by camera manufacturers and platforms, and it can be defeated by re-recording or by stripping metadata \citep{corsi2024synthetic}.

In the health domain, where AI-generated text received the highest threat ratings of the four modalities (M = 5.80), respondents suggested greater regulatory attention toward labeling AI-generated health content and improving medical fact-checking infrastructure \citep{deangelis2023chatgpt}. The World Health Organization\textquotesingle s infodemic-management framework \citep{who2020infodemic} offers an operational template: it pairs active listening for circulating health claims with rapid authoritative rebuttal and amplification through trusted local messengers. It was developed before current large language models, but because it targets the circulation of false health claims rather than their means of production, it transfers to AI-generated text with little modification. Experts gave audio deepfakes the highest average threat ratings in the financial domain (\emph{M} = 5.65), aligning with growing evidence on voice-cloning fraud. In 2024, a Hong Kong finance worker was deceived into transferring \$25 million through a deepfake video call \citep{chen2024finance}. The U.S. Treasury's Financial Crimes Enforcement Network has since issued a sector-wide alert on deepfake-enabled fraud against financial institutions \citep{fincen2024alert}---an early example of the type of domain-specific guidance highlighted by many respondents. The expert responses suggest that voice-authentication standards for high-value transactions deserve increased attention, and they raise questions about the adequacy of existing know-your-customer (KYC) protocols in an environment where a live voice or video is no longer reliable evidence of identity \citep{chesney2019deepfakes}.

\subsection{A contested, not polarized, mitigation landscape}\label{a-contested-not-polarized-mitigation-landscape}

Respondents rated five mitigation strategies: government regulation, digital watermarking, media literacy, platform enforcement, and technical detection. Likert effectiveness ratings for all five strategies are right-skewed and unimodal: the share of respondents rating a strategy 5 or higher on the 7-point scale ranged from 63\% (technical detection) to 74\% (government regulation), with no bimodal distribution. The two question formats point in different directions. On the Likert ratings, every strategy is broadly endorsed. On the forced-choice ranking, no strategy commands a majority and the same strategy can appear at both ends: government regulation drew 30\% of ``most effective'' votes but also 15\% of ``least effective'' votes, and media literacy 26\% against 24\%. Experts agree that each strategy can contribute; they disagree on which should come first. The landscape is contested rather than polarized, mirroring a recent \emph{HKS Misinformation Review} survey on generative AI in Europe \citep{weikmann2026samepage}.

Psychological inoculation deserves a more prominent place in this priority debate than our respondents gave it; it was not among the five strategies we asked about, and no respondent volunteered it. Pre-exposing audiences to weakened forms of misinformation techniques builds resistance that persists across topics and time horizons \citep{vanderlinden2023foolproof}, and game-based inoculation has shown durable real-world effects at scale \citep{roozenbeek2024psychology}. Detection tools face a perpetual arms race with generative AI models \citep{hoq2025feedback,murphy2023scoping,rana2022deepfake}. Direct quotations below are attributed by name only to the 13 respondents who explicitly consented to such attribution; all other respondents are described generically. One respondent, the researcher and digital artist publishing as Merzmensch, captured the urgency: ``Literacy! Literacy! Literacy! From the first classes in school.'' Another respondent, technology professional Alberto Lobato Diogo, underscored the layered character of any realistic response: ``GenAI based on LLMs are mathematically impossible to have full-proof mitigation systems, so we need a combination of all of the above.''

\subsection{Public-awareness tools reach the wrong audiences}\label{public-awareness-tools-reach-the-wrong-audiences}

Experts rated public-awareness tools as moderately effective (\emph{M} = 4.43), and one third identified the same central limitation: these tools reach mainly audiences that are already technically confident. \citet{guo2025susceptible} report the same asymmetry from a different angle, finding that the populations most vulnerable to AI-generated disinformation are also the least likely to adopt technical countermeasures. Our respondents\textquotesingle{} assessment therefore suggests that effective mitigation may need to extend beyond tool development to distribution, accessibility, and integration into platforms where vulnerable users already consume information. One respondent, Katerina Sedova of the Atlantic Council, warned: ``As chatbots become ubiquitous, get intertwined with human lives, and even invite real emotional connection and dependence from humans, it will be critical to ensure that these mediums are not weaponized.''

\emph{Toward a layered, domain-weighted response}

Because no single strategy commands consensus while each is broadly seen as workable, the responses point toward combination rather than selection. Grouping the five rated strategies by the mechanism through which they act yields four pillars: provenance (digital watermarking and technical detection, which act on the artifact), audience-side literacy (media literacy, which acts on the recipient), statutory regulation (government regulation, which acts on the producer), and platform enforcement (which acts on distribution). The number four here is coincidental and carries no relation to the four modalities or the four domains.

The pillars are not weighted equally across domains. Our respondents\textquotesingle{} domain-by-modality ratings suggest different entry points: provenance first for political video, where authenticity of the artifact is the contested question; labeling and medical fact-checking for health text, where the claim rather than the medium carries the harm; voice authentication and KYC modernization for financial audio, where the attack targets an identity check; and accessibility-first literacy for the social domain, where harm is diffuse and no single chokepoint exists. Appendix Table C4 sets out this priority matrix. It is a starting point derived from perceptions, not a validated framework, and we present it as a hypothesis for testing.

\section{Findings}\label{findings}

\subsection{Finding 1: Experts rate video deepfakes as the most threatening modality overall. }\label{finding-1-experts-rate-video-deepfakes-as-the-most-threatening-modality-overall.}

\begin{figure}[htbp]
\centering
\includegraphics[width=\linewidth,keepaspectratio]{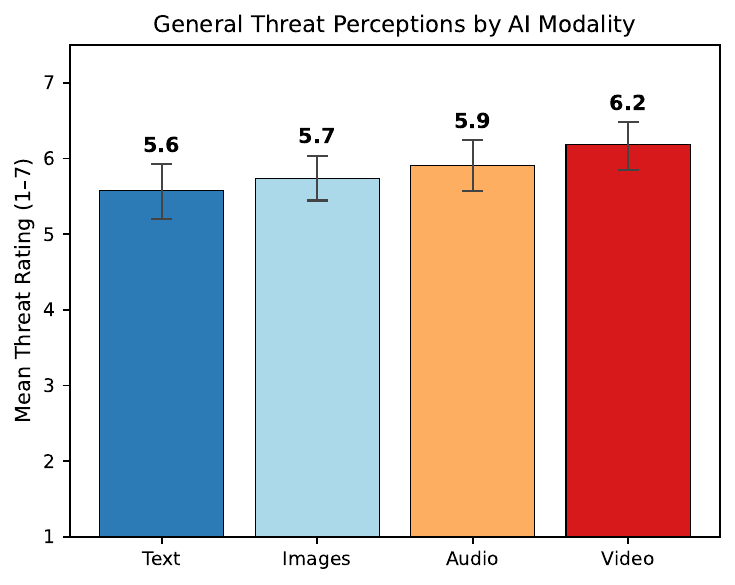}
\caption{Modality threat perceptions across all domains (7-point scale, N = 54). Error bars represent 95\% bootstrap confidence intervals (10,000 iterations).}
\end{figure}

Throughout the Findings, ratings use a 7-point scale (1 = not threatening; 7 = extremely threatening) \emph{MSD}. Across all domains, video deepfakes received the highest average threat rating (\emph{M} = 6.19, \emph{SD} = 1.19), followed by audio (\emph{M} = 5.91, \emph{SD} = 1.25), images (\emph{M} = 5.74, \emph{SD} = 1.13), and text (\emph{M} = 5.57, \emph{SD} = 1.36). This pattern is broadly consistent with previous work suggesting that synthetic content in higher-fidelity modalities carries greater persuasive force, and that untrained observers find fabricated video and audio harder to identify as synthetic than fabricated text \citep{corsi2024synthetic,guo2025susceptible}.

\subsection{Finding 2: Perceived modality threats vary systematically across domains. }\label{finding-2-perceived-modality-threats-vary-systematically-across-domains.}

\begin{figure}[htbp]
\centering
\includegraphics[width=\linewidth,keepaspectratio]{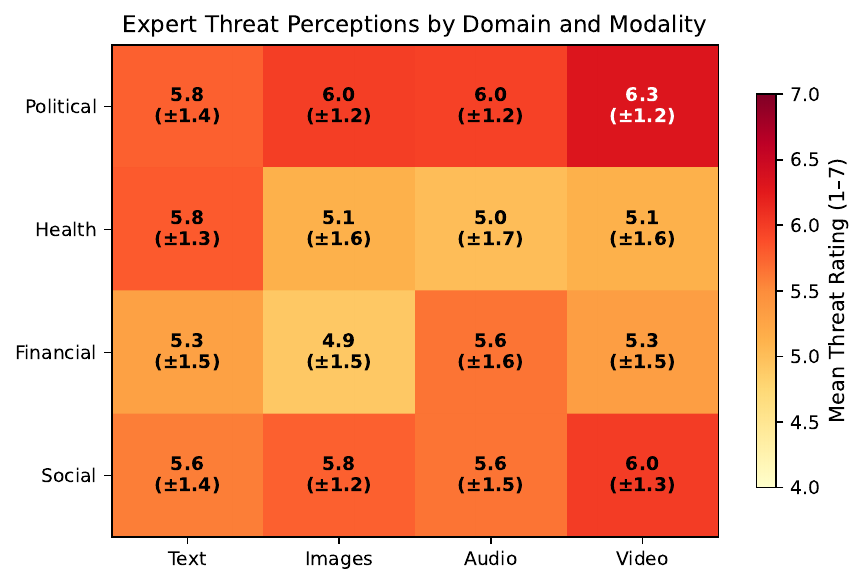}
\caption{Domain × Modality threat perception heatmap~(N = 54). Cell values show M (±SD). Higher values indicate greater perceived threat.}
\end{figure}

Within this sample, the political domain received the highest average threat ratings, with deepfake video peaking at \emph{M} = 6.31 (\emph{SD} = 1.15). The confidence intervals political-domain video and audio overlap, however. In plain terms: with 54 respondents, the difference between these two ratings is small enough that it could plausibly be a product of who happened to answer the survey rather than a real difference in expert opinion. The ordering should be read as suggestive, not established.

The social domain followed a similar pattern (video \emph{M} = 6.00; images \emph{M} = 5.76). The health domain showed a different pattern: text was rated as the most threatening (\emph{M} = 5.80, \emph{SD} = 1.32), while images (\emph{M} = 5.13), audio (\emph{M} = 5.02), and video (\emph{M} = 5.13) clustered lower. However, confidence intervals overlap across all four health-domain modalities, so this pattern should be treated as suggestive rather than definitive. One possible explanation is that health misinformation reaches audiences mainly in written form, from fabricated studies to AI-generated health advice, so experts see text as the more likely route to harm in this domain \citep{deangelis2023chatgpt}.

\begin{figure}[htbp]
\centering
\includegraphics[width=\linewidth,keepaspectratio]{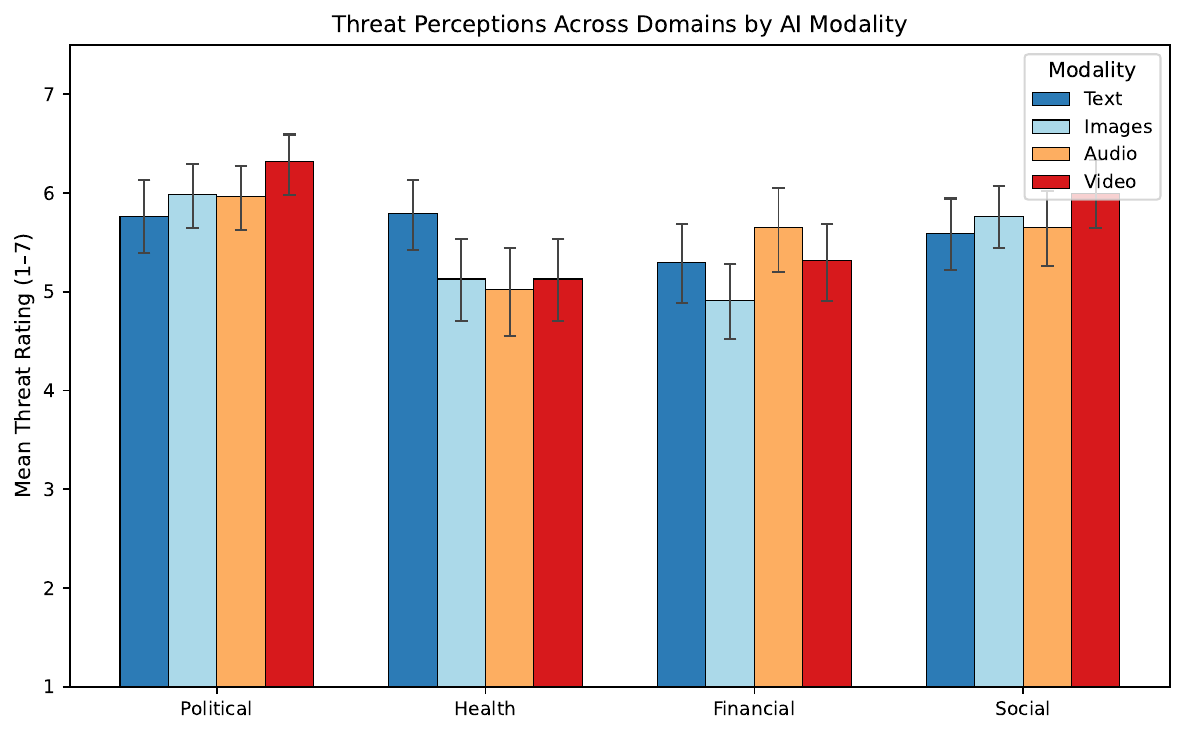}
\caption{Threat perceptions across domains by AI modality (N = 54). Error bars represent 95\% bootstrap confidence intervals (10,000 iterations). Substantial overlap indicates that most cross-domain and cross-modality differences are not statistically distinguishable at this sample size.}
\end{figure}

In the financial domain, audio received the highest mean rating (\emph{M} = 5.65, \emph{SD} = 1.62), though confidence intervals do not clearly separate it from video or text.

\subsection{Finding 3: Experts identify voice-cloning fraud as the most urgent risk. }\label{finding-3-experts-identify-voice-cloning-fraud-as-the-most-urgent-risk.}

\begin{figure}[htbp]
\centering
\includegraphics[width=\linewidth,keepaspectratio]{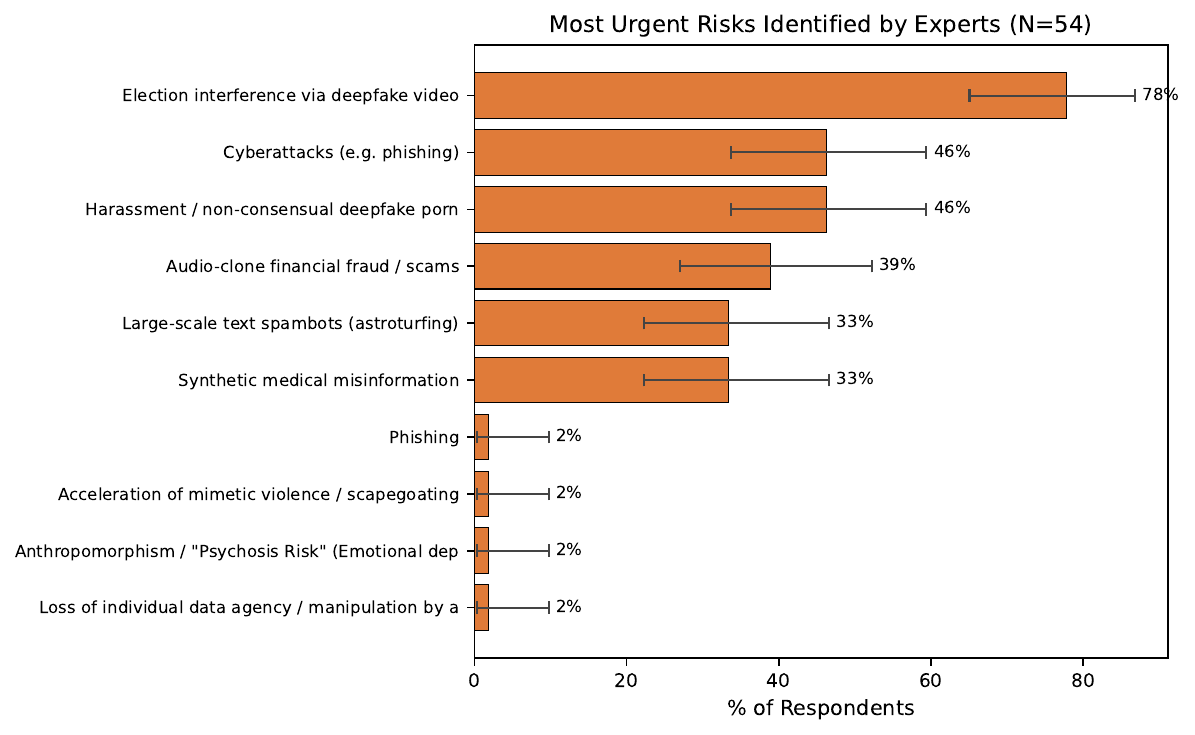}
\caption{Most urgent risks identified by experts~(N = 54). Error bars represent 95\% Wilson score confidence intervals for binomial proportions.}
\end{figure}

Election interference via deepfake video was identified as the most urgent risk by 78\% of respondents, consistent with widespread public concern documented in recent surveys of U.S. voters \citep{yan2025origin}. The next tier comprised cyberattacks (46\%) and non-consensual deepfake imagery (46\%), though Wilson confidence intervals for these proportions are wide (approximately 33\%--60\%), indicating substantial uncertainty in the precise ranking below election interference.

\subsection{Finding 4: Experts view no single mitigation strategy as universally effective. }\label{finding-4-experts-view-no-single-mitigation-strategy-as-universally-effective.}

\begin{figure}[htbp]
\centering
\includegraphics[width=\linewidth,keepaspectratio]{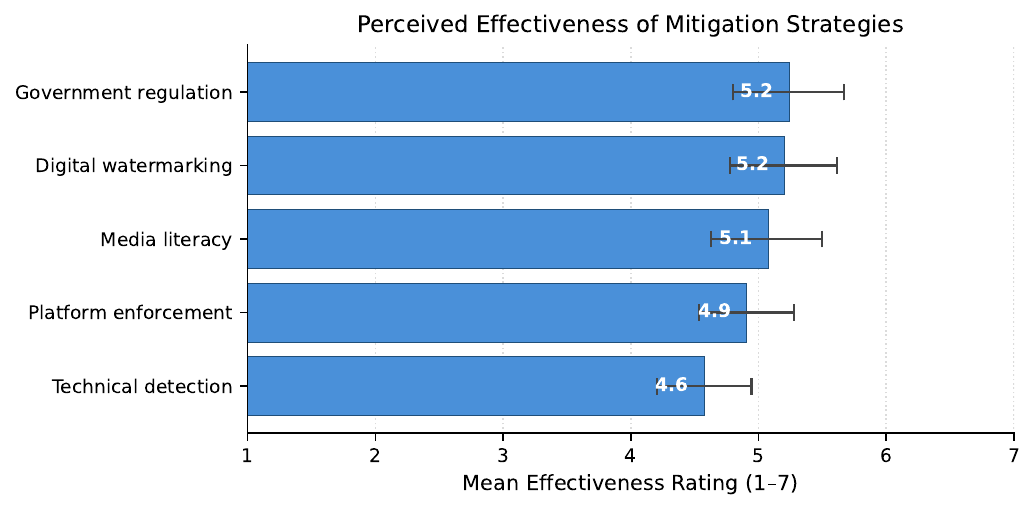}
\caption{Mitigation strategy effectiveness ratings (7-point scale, N = 54). Error bars represent 95\% bootstrap confidence intervals (10,000 iterations).}
\end{figure}

On a 7-point effectiveness scale, government regulation received the highest mean rating (\emph{M} = 5.24, \emph{SD} = 1.65), followed by digital watermarking (\emph{M} = 5.20), media literacy (\emph{M} = 5.07), platform enforcement (\emph{M} = 4.91), and technical detection (\emph{M} = 4.57). Bootstrap confidence intervals overlap for the top four strategies, indicating that the differences among them are not statistically distinguishable; only technical detection falls somewhat below the others. All five distributions are right-skewed and unimodal (63--74\% of respondents at ≥ 5; see Appendix Table C2). In a separate item, respondents ranked the same five strategies from 1 (most effective) to 5 (least effective). Mean ranks fell in a narrow band from 2.98 to 3.28, close to the midpoint of 3 that would result from no agreement at all. This aligns with prior expert survey research showing persistent disagreement about optimal responses to misinformation \citep{altay2023survey}.

\section{Methods}\label{methods}

This study is exploratory. We did not pre-register hypotheses; we sought a broad descriptive map of expert threat assessments across modalities, domains, and mitigation strategies. The present manuscript reports preliminary findings from this dataset.

Going in, we held three loose expectations grounded in the existing literature: (a) higher-fidelity modalities (video, audio) would dominate threat ratings, with video especially salient in the political domain \citep{corsi2024synthetic}; (b) audio threats would cluster in the financial domain, given the rise of voice-cloning fraud \citep{chen2024finance}; and (c) experts would split into recognizable camps along a tech-fix-versus-regulation axis \citep{altay2023survey}. The data confirmed (a) and (b) but contradicted (c): rather than camp-formation, we observed broad agreement that each strategy can work alongside disagreement on which deserves priority. We treat this contrast between expectation and finding as a hypothesis source for confirmatory follow-up rather than as a tested claim.

We conducted a structured online survey between July 2025 and March 2026. Eligible participants were professionals or scholars with demonstrated recent activity on AI-driven disinformation; eligibility was operationalized by requiring respondents to provide a link to a recent publication, project, or public engagement on the topic. The link served as a soft validation; no formal post-response screening or exclusion was applied, and all 54 valid responses are retained in the analyses.

Recruitment combined targeted expert recruitment with snowball extension. We identified 516 candidates through systematic Google Scholar, LinkedIn, Mastodon, and Bluesky searches, supplemented by authors of recent publications cited in our broader research program. We directly contacted 454 candidates via individualized messages (email 47\%, LinkedIn 21\%, Mastodon DM 8\%, Bluesky 4\%, mixed/other 20\%); recipients were invited to forward the survey to relevant colleagues. The questionnaire (54 items, mostly 1-to-7 rating scales) is reproduced in Appendix A; the recoding scheme for free-text role descriptions is documented in Appendix B.

We received 54 valid responses (response rate 11.9\% from over 454 directly contacted; the absolute number of forwards via snowball is unknown). The equal number of items (54) and respondents (54) is coincidental. Recoding the open-text role field yielded seven coherent categories: AI/ML researchers and developers (39\%); disinformation, fact-checking, journalism, or media research (26\%); other practitioners and academics (13\%); cybersecurity, defense, and threat-intelligence (6\%); policymakers and regulators (6\%); provenance, standards, and infrastructure professionals (6\%); and ethics, legal, and governance (6\%). Respondents had a median of 17 years of experience and high self-rated familiarity with both large language models (\emph{M} = 4.89/7) and deepfake technology (\emph{M} = 4.54/7). Respondents were distributed across five geographic areas: European Union (n = 23, 43\%), North America (n = 19, 35\%), non-EU Europe including the United Kingdom (n = 7, 13\%), Asia-Pacific (n = 4, 7\%, comprising East and Southeast Asia, South and Central Asia, and Oceania), and one respondent (2\%) who selected ``Global,'' an option offered for experts whose work is not tied to a single region. No respondent was based in Africa or in Latin America, although both were available options. Sample characteristics by category appear in Appendix Table C1.

Participants chose one of three publication-handling preferences at the end of the survey: full anonymity (\emph{n} = 27, 50.0\%), name and affiliation in Acknowledgments (\emph{n} = 14, 25.9\%), or attribution of specific quotes by name in addition to acknowledgment (\emph{n} = 13, 24.1\%). All four respondents quoted by name in this manuscript fall in the third group; ``Merzmensch'' is a long-standing artistic identity under which the respondent has published creative and scholarly work and was supplied as the preferred attribution.

Quantitative data were analyzed descriptively (means, standard deviations, frequencies). Consequently, the reported differences should be interpreted as descriptive patterns rather than evidence of statistically significant differences between modalities or domains. Given the exploratory design and sample size, we report descriptive statistics rather than inferential tests. To convey estimation uncertainty, figures display 95\% bootstrap confidence intervals (10,000 iterations with a fixed random seed for reproducibility) for mean ratings, and 95\% Wilson score confidence intervals for binomial proportions.

\subsection{Limitations}\label{limitations}

The sample (\emph{N} = 54) is purposive, not representative, and skews toward Western, technically proficient experts: 91\% of respondents are based in the European Union, North America, or non-EU Europe, with the Global South substantially underrepresented. Self-selection bias may inflate threat perceptions among respondents already concerned about AI-driven disinformation. The survey measures perceptions rather than observed impacts, and the rapidly evolving capabilities of generative AI may shift expert assessments substantially within months. Sub-domain comparisons are descriptive only and underpowered for inferential testing.

Because the sample contains a comparatively large proportion of AI researchers and disinformation specialists, the observed priorities likely reflect expert perspectives on AI-related risks rather than broader societal perceptions. This composition may have influenced which domains were viewed as most concerning.

All respondents were 18 or older and provided informed consent prior to participation. The consent form, displayed at the start of the survey, described the study purpose, voluntary nature of participation, anticipated time commitment, data-handling procedures, and the three publication-handling options described above. The survey did not collect special-category personal data and was conducted in accordance with the EU General Data Protection Regulation. The study protocol was reviewed against the data-minimization and lawful-basis requirements of the EU General Data Protection Regulation (consent under Art. 6(1)(a) GDPR). Direct quotations are attributed by name only for the 13 respondents who explicitly opted in to such attribution.

\bibliographystyle{abbrvnat}
\bibliography{references}

\begin{thebibliography}{17}
\providecommand{\natexlab}[1]{#1}
\providecommand{\url}[1]{\texttt{#1}}
\expandafter\ifx\csname urlstyle\endcsname\relax
  \providecommand{\doi}[1]{doi: #1}\else
  \providecommand{\doi}{doi: \begingroup \urlstyle{rm}\Url}\fi

\bibitem[Altay et~al.(2023)Altay, Berriche, Heuer, Farkas, and
  Rathje]{altay2023survey}
S.~Altay, M.~Berriche, H.~Heuer, J.~Farkas, and S.~Rathje.
\newblock A survey of expert views on misinformation: Definitions,
  determinants, solutions, and future of the field.
\newblock \emph{Harvard Kennedy School Misinformation Review}, 4\penalty0 (4),
  2023.
\newblock \doi{10.37016/mr-2020-119}.

\bibitem[Bommasani et~al.(2024)Bommasani, Klyman, Kapoor, Longpre, Xiong,
  Maslej, and Liang]{bommasani2024fmti}
R.~Bommasani, K.~Klyman, S.~Kapoor, S.~Longpre, B.~Xiong, N.~Maslej, and
  P.~Liang.
\newblock The 2024 foundation model transparency index, 2024.
\newblock URL \url{https://arxiv.org/abs/2407.12929}.

\bibitem[Chen and Magramo(2024)]{chen2024finance}
H.~Chen and K.~Magramo.
\newblock Finance worker pays out \$25 million after video call with deepfake
  ``chief financial officer'', February 2024.
\newblock URL
  \url{https://www.cnn.com/2024/02/04/asia/deepfake-cfo-scam-hong-kong-intl-hnk/index.html}.
\newblock CNN.

\bibitem[Chesney and Citron(2019)]{chesney2019deepfakes}
R.~Chesney and D.~K. Citron.
\newblock Deep fakes: A looming challenge for privacy, democracy, and national
  security.
\newblock \emph{California Law Review}, 107\penalty0 (6):\penalty0 1753--1820,
  2019.
\newblock \doi{10.2139/ssrn.3213954}.

\bibitem[Corsi et~al.(2024)Corsi, Marino, and Wong]{corsi2024synthetic}
G.~Corsi, B.~Marino, and W.~Wong.
\newblock The spread of synthetic media on {X}.
\newblock \emph{Harvard Kennedy School Misinformation Review}, 5\penalty0 (3),
  2024.
\newblock \doi{10.37016/mr-2020-140}.

\bibitem[De~Angelis et~al.(2023)De~Angelis, Baglivo, Arzilli, Privitera,
  Ferragina, Tozzi, and Rizzo]{deangelis2023chatgpt}
L.~De~Angelis, F.~Baglivo, G.~Arzilli, G.~P. Privitera, P.~Ferragina, A.~E.
  Tozzi, and C.~Rizzo.
\newblock {ChatGPT} and the rise of large language models: The new {AI}-driven
  infodemic threat in public health.
\newblock \emph{Frontiers in Public Health}, 11:\penalty0 1166120, 2023.
\newblock \doi{10.3389/fpubh.2023.1166120}.

\bibitem[{European Parliament}(2024)]{europarl2024aiact}
{European Parliament}.
\newblock Regulation (eu) 2024/1689 laying down harmonised rules on artificial
  intelligence (ai act), 2024.
\newblock URL
  \url{https://eur-lex.europa.eu/legal-content/EN/TXT/?uri=OJ:L_202401689}.

\bibitem[{Financial Crimes Enforcement Network}(2024)]{fincen2024alert}
{Financial Crimes Enforcement Network}.
\newblock {FinCEN} alert on fraud schemes involving deepfake media targeting
  financial institutions (fin-2024-alert004), November 2024.
\newblock URL
  \url{https://www.fincen.gov/sites/default/files/shared/FinCEN-Alert-DeepFakes-Alert508FINAL.pdf}.
\newblock U.S. Department of the Treasury.

\bibitem[Guo et~al.(2025)Guo, Zhong, and Hu]{guo2025susceptible}
S.~Guo, Y.~Zhong, and X.~Hu.
\newblock People are more susceptible to misinformation with realistic
  {AI}-synthesized images that provide strong evidence to headlines.
\newblock \emph{Harvard Kennedy School Misinformation Review}, 6\penalty0 (6),
  2025.
\newblock \doi{10.37016/mr-2020-189}.

\bibitem[Hoq et~al.(2025)Hoq, Facciani, and Weninger]{hoq2025feedback}
A.~Hoq, M.~J. Facciani, and T.~Weninger.
\newblock Feedback and education improve human detection of image manipulation
  on social media.
\newblock \emph{Harvard Kennedy School Misinformation Review}, 6\penalty0 (2),
  2025.
\newblock \doi{10.37016/mr-2020-175}.

\bibitem[Murphy et~al.(2023)Murphy, de~Saint~Laurent, Reynolds, Aftab, Hegarty,
  Sun, and Greene]{murphy2023scoping}
G.~Murphy, C.~de~Saint~Laurent, M.~Reynolds, O.~Aftab, K.~Hegarty, Y.~Sun, and
  C.~M. Greene.
\newblock What do we study when we study misinformation? a scoping review of
  experimental research (2016--2022).
\newblock \emph{Harvard Kennedy School Misinformation Review}, 4\penalty0 (6),
  2023.
\newblock \doi{10.37016/mr-2020-130}.

\bibitem[Rana et~al.(2022)Rana, Nobi, Murali, and Sung]{rana2022deepfake}
M.~S. Rana, M.~N. Nobi, B.~Murali, and A.~H. Sung.
\newblock Deepfake detection: A systematic literature review.
\newblock \emph{IEEE Access}, 10:\penalty0 25494--25513, 2022.
\newblock \doi{10.1109/ACCESS.2022.3154404}.

\bibitem[Roozenbeek and van~der Linden(2024)]{roozenbeek2024psychology}
J.~Roozenbeek and S.~van~der Linden.
\newblock \emph{The psychology of misinformation}.
\newblock Cambridge University Press, 2024.
\newblock \doi{10.1017/9781009214414}.

\bibitem[van~der Linden(2023)]{vanderlinden2023foolproof}
S.~van~der Linden.
\newblock \emph{Foolproof: Why misinformation infects our minds and how to
  build immunity}.
\newblock W. W. Norton, 2023.

\bibitem[Weikmann et~al.(2026)]{weikmann2026samepage}
T.~Weikmann et~al.
\newblock On the same page? experts are mostly, but not always aligned on
  generative {AI} and misinformation in {Europe}.
\newblock \emph{Harvard Kennedy School Misinformation Review}, 2026.
\newblock \doi{10.37016/mr-2020-196}.

\bibitem[{World Health Organization}(2020)]{who2020infodemic}
{World Health Organization}.
\newblock Managing the {COVID-19} infodemic: Promoting healthy behaviours and
  mitigating the harm from misinformation and disinformation.
\newblock
  \url{https://www.who.int/news/item/23-09-2020-managing-the-covid-19-infodemic-promoting-healthy-behaviours-and-mitigating-the-harm-from-misinformation-and-disinformation},
  2020.

\bibitem[Yan et~al.(2025)]{yan2025origin}
H.~Yan et~al.
\newblock The origin of public concerns over {AI} supercharging misinformation.
\newblock \emph{Harvard Kennedy School Misinformation Review}, 2025.
\newblock \doi{10.37016/mr-2020-171}.

\end{thebibliography}

\section{Acknowledgements }\label{acknowledgements}

We gratefully acknowledge the 54 experts who contributed their time and expertise to this survey. We especially thank the following respondents who consented to be named: Marc-Oliver Pahl (IMT Atlantique), Gürkan Solmaz (ACM Europe TPC), Gerhard Schimpf, Volker Klaeren, Merzmensch, Arne Stenmanns, Melanie Siegel (Darmstadt University of Applied Sciences), Filippo Menczer (Observatory on Social Media, Indiana University), Jeff Jarvis (CUNY), Stephan Lewandowsky (University of Bristol), Emilio Ferrara (University of Southern California), Timothy R. Levine (University of Oklahoma), Jesper Strömbäck (University of Gothenburg), Giovanni Spitale (University of Zurich), Eelco Herder (Utrecht University), Alberto Lobato Diogo, Sam Stockwell (The Alan Turing Institute), Evan M. Williams (Carnegie Mellon University), Peter Carragher (Carnegie Mellon University), Manju Rose Mathews (Christ Nagar College), Katerina Sedova (Atlantic Council), William Kempster (ar.io network), Scott Perry (Digital Governance Institute), Doug Finke, Roberto V. Zicari (Z-Inspection® Initiative), Jason Potel (Goldsmiths, University of London), and Muhammad Zubair (Fortex Solutions).

\section{Authorship}\label{authorship}

A. L. conceived and designed the study, developed the survey instrument, conducted data collection, performed the analysis, and wrote the manuscript. M. K. and M.-O. P. provided supervision and critical feedback on the study design and manuscript. All authors reviewed and approved the final version.

\section{Funding}\label{funding}

No funding has been received to conduct this research.

\section{Competing interests}\label{competing-interests}

The authors declare no competing interests.

\section{Ethics}\label{ethics}

This study was conducted in accordance with ethical guidelines for human subjects research. All participants provided informed consent prior to completing the survey, confirming they were 18 years or older and understood the purpose, voluntary nature, and data handling procedures of the study. Participants could withdraw at any time without consequences. The survey collected professional opinions on AI-generated disinformation threats and mitigation strategies; no sensitive personal data, medical information, or vulnerable populations were involved. Respondents self-selected their attribution preference, choosing whether to remain anonymous, be acknowledged by name, or allow direct quotation. Ethnicity and gender were not collected, as they were not relevant to the research questions. The study was reviewed and approved by the supervising institution. Direct quotations are attributed by name only for the 13 respondents who explicitly opted in to such attribution. One of these attributions uses a pseudonymous professional identity (``Merzmensch''), an established public byline that we verified in direct correspondence with the respondent rather than through institutional affiliation.

\section{Copyright}\label{copyright}

This is an open access article distributed under the terms of the \href{https://creativecommons.org/licenses/by/4.0/}{Creative Commons Attribution License}, which permits unrestricted use, distribution, and reproduction in any medium, provided that the original author and source are properly credited.

\section{Data availability}\label{data-availability}

All materials needed to replicate this study are available via the Harvard Dataverse: \url{https://doi.org/10.7910/DVN/BXO2QA}. The deposit includes: (i) the verbatim survey instrument as administered (also reproduced in Appendix A below), (ii) the aggregated, de-identified response tables underlying all figures and supplementary tables, (iii) the analysis code used to produce them, and (iv) the recoding scheme for the open-text role field (Appendix B). Individual-level free-text responses are withheld within IRB restrictions and in keeping with respondents' anonymity preferences (50.0\% of participants opted for full anonymity); only aggregated counts and consented quotations are released.

\section{Appendix A: Survey instrument}\label{appendix-a-survey-instrument}

The instrument was administered in English via Google Forms between July 3, 2025, and March 15, 2026. It is organized in nine sections plus consent, future-outlook, and attribution blocks. Several items use a modality- or strategy-keyed grid (e.g., the four threat-domain matrices each elicit 16 modality~×~domain ratings), which is how the 34 numbered questions yield the 54 substantive data points per respondent referenced in the main text. Items marked * were required; all others were optional. Likert-type response options are stated once per scale. The full LaTeX source of the instrument is included in the Harvard Dataverse deposit (see Data Availability section); the structured summary below preserves item wording, ordering, and response options.

\textbf{A.1 Consent \& data protection (1 item).} GDPR notice naming the data controller (Alexander Loth, Frankfurt UAS), purpose, lawful basis (explicit consent), retention, and participant rights (withdrawal, access, rectification, erasure, complaint to the Hessian Data Protection Commissioner). \emph{Q1*} --- ``I confirm I am 18 years or older and I have read, understood, and agree to the consent terms outlined above.'' --- \emph{Yes / No} (gating).

\textbf{A.2 Section 1/9 --- Screening \& expert group (2 items).} \emph{Q2*} Primary professional role --- \emph{single choice:} AI researcher/ML engineer; disinformation/fact-checking professional; journalist covering tech or politics; policymaker/regulator; ethicist/legal scholar; other (free text). Recoded post hoc into seven categories (see Appendix B). \emph{Q3*} Link to a recent activity (recent paper, organizational role, profile link) --- \emph{free text.}

\textbf{A.3 Section 2/9 --- Demographics \& baseline knowledge (4 items).} \emph{Q5*} Years of professional experience --- \emph{numeric.} \emph{Q6*} Region of primary professional activity --- \emph{single choice:} European Union; Europe (non-EU, incl.~UK); North America (USA \& Canada); East \& Southeast Asia; South \& Central Asia; Latin America \& Caribbean; MENA; Sub-Saharan Africa; Oceania; other. \emph{Q7*} Self-rated technical understanding of Large Language Models --- \emph{1--7 (Novice--Expert).} \emph{Q8*} Self-rated familiarity with deepfake video technology --- \emph{1--7 (Novice--Expert).}

\textbf{A.4 Section 3/9 --- Perceived threats, modality only (1 item).} \emph{Q9.} ``How threatening is each modality for spreading disinformation in general?'' --- \emph{1--7 grid (Not a threat--Severe threat) for:} AI-generated text, images, audio (voice cloning), video (deepfakes).

\textbf{A.5 Section 4/9 --- Perceived threats, modality × domain matrix (4 rated items + 4 optional comments).} Stem (repeated per domain): ``Threats in the \textbf{\{DOMAIN\}} Domain (1~=~Not a threat, 7~=~Severe threat).'' Domains: \textbf{Political} (Q10), \textbf{Health} (Q12), \textbf{Financial} (Q14), \textbf{Social} (Q16). Each is a 4-row × 7-column grid (text/images/audio/video). Each domain is followed by an \emph{optional} free-text comment item (Q11, Q13, Q15, Q17): ``Which specific points concerning the \{DOMAIN\} Domain would you like to share with us?''

\textbf{A.6 Section 5/9 --- Key risks, top-three (1 item + 1 optional).} \emph{Q18*} Select up to three risks judged most urgent --- \emph{multi-select (max 3):} election interference via deepfake video; large-scale text spambots (astroturfing); audio-clone financial fraud/scams; synthetic medical misinformation; harassment/non-consensual deepfake porn; cyberattacks (e.g.,~phishing); other (free text). \emph{Q19.} In 1--2 sentences, why these worry you most --- \emph{optional, free text.}

\textbf{A.7 Section 6/9 --- Mitigation strategies, ratings (1 item + 1 optional).} \emph{Q20.} ``Rate the effectiveness of the following mitigation strategies'' --- \emph{1--7 grid (Not effective--Very effective) for:} technical detection tools; digital watermarking/provenance standards; government regulation (e.g., AI Act, DSA); media/public literacy programs; platform enforcement policies. \emph{Q21.} Additional information on mitigation strategies --- \emph{optional, free text.}

\textbf{A.8 Section 7/9 --- Mitigation strategies, forced ranking (1 item).} \emph{Q22.} Rank the five strategies above from 1 (most effective) to 5 (least effective) --- \emph{full ranking, no ties.}

\textbf{A.9 Section 8/9 --- Best--worst scaling (3 items).} \emph{Q23*} Which strategy is \textbf{most} effective? --- \emph{single choice from the same five strategies.} \emph{Q24*} Which strategy is \textbf{least} effective? --- \emph{single choice from the same five strategies.} \emph{Q25.} Single biggest obstacle (technical, political, or economic) blocking adoption of your top-ranked strategy --- \emph{optional, free text.}

\textbf{A.10 Section 9/9 --- Public-literacy tools (4 items).} A short methodology note describes a representative tool, JudgeGPT (\url{https://github.com/aloth/JudgeGPT}), in which respondents view five short news fragments and make two slider-based judgments per fragment (origin: human vs.~AI; veracity: legitimate vs.~fake), with accuracy feedback and badges. \emph{Q26.} Effectiveness of such public-awareness tools for mitigating AI-generated fake news --- \emph{1--7 (Not Effective--Very Effective).} \emph{Q27*} Single greatest limitation --- \emph{single choice:} easily bypassed by adversarial content; reaches only tech-savvy audiences; impact fades once novelty wears off; resource-intensive to maintain; other (free text). \emph{Q28.} Main strength of such tools --- \emph{optional, free text.} \emph{Q29.} One concrete feature or program that would make public-resilience tools more effective --- \emph{optional, free text.}

\textbf{A.11 Future outlook \& final comments (2 items).} \emph{Q30.} One emerging or currently underestimated risk at the AI--disinformation frontier on a five-year horizon --- \emph{optional, free text.} \emph{Q31.} Final thoughts or feedback on the survey --- \emph{optional, free text.}

\textbf{A.12 Attribution \& final submission (4 items).} \emph{Q32*} Handling of contribution in the final publication --- \emph{single choice:} keep responses fully anonymous; list name and affiliation in the Acknowledgments; in addition, attribute specific quotes or ideas by name. \emph{Q33*} Full name --- \emph{free text.} \emph{Q34*} Affiliation --- \emph{free text.} \emph{Q35.} Link to preferred professional/academic profile --- \emph{optional, free text.}

\section{Appendix B: Recoding scheme for professional roles}\label{appendix-b-recoding-scheme-for-professional-roles}

The survey collected primary professional role as a free-text field. Thirty-two distinct strings were submitted. We recoded these into seven coherent categories using the rules below, applied conservatively to preserve respondent self-description.

\emph{\textbf{Table B1.} Recoding scheme mapping open-text professional role descriptions to seven coherent categories (}N \emph{= 54).}

{\def\LTcaptype{none} 
\begin{longtable}[]{@{}
  >{\raggedright\arraybackslash}p{(\linewidth - 2\tabcolsep) * \real{0.5000}}
  >{\raggedright\arraybackslash}p{(\linewidth - 2\tabcolsep) * \real{0.5000}}@{}}
\toprule\noalign{}
\begin{minipage}[b]{\linewidth}\raggedright
\textbf{Category (n)}
\end{minipage} & \begin{minipage}[b]{\linewidth}\raggedright
\textbf{Includes raw strings such as}
\end{minipage} \\
\midrule\noalign{}
\endhead
\bottomrule\noalign{}
\endlastfoot
AI/ML researcher or developer (21) & ``AI researcher/ML engineer,'' ``Software developer, deep in the AI space,'' ``AI trainer,'' ``Professor for semantic technologies and computer science,'' ``Economist working w AI,'' ``Trustworthy AI co-creation and assessment,'' ``Product management,'' ``IT architect'' \\
Disinformation, fact-checking, journalism, or media research (14) & ``Disinformation/fact-checking professional,'' ``Journalist covering tech or politics,'' ``Misinformation academic researcher,'' ``Communications researcher,'' ``Journalism scholar researching the impact of AI,'' ``Media psychology researcher,'' ``University researcher of information ecosystems,'' ``Researcher on disinformation, AI, etc.,'' ``Deception researcher and author of TDT'' \\
Other practitioners \& academics (7) & ``Teacher,'' ``Private person (IT consultant),'' ``Culture and art history researcher,'' ``Investor/venture capitalist (focus on AI and defense tech),'' ``Academic,'' ``Technical training'' \\
Cybersecurity/defense/threat intelligence (3) & ``Professor cybersecurity,'' ``Subject matter expert and researcher of AI-enabled threats and opportunities in cyber, information environment, and defense technology'' \\
Policymaker/regulator (3) & ``Policymaker/regulator involved with AI, media or digital-services policy,'' ``Member of the ACM Europe Technology Policy Committee'' \\
Provenance/standards/infrastructure (3) & ``Infrastructure provider supplying verifiable data storage and access,'' ``Advocate for adoption of C2PA provenance technologies,'' ``Technologist/web standards advocate (focus on decentralization and data sovereignty)'' \\
Ethics, legal, or governance (3) & ``Ethicist/legal scholar working on AI or deepfakes,'' ``Governance consultant and C2PA conformance program administrator'' \\
\end{longtable}
}

\emph{Note: Unique role strings: 32. All recoded; none discarded.}

\textbf{Appendix C: Supplementary statistical tables}

\emph{\textbf{Table C1.} Sample characteristics of the expert respondents (N = 54).}

{\def\LTcaptype{none} 
\begin{longtable}[]{@{}
  >{\raggedright\arraybackslash}p{(\linewidth - 2\tabcolsep) * \real{0.6651}}
  >{\centering\arraybackslash}p{(\linewidth - 2\tabcolsep) * \real{0.1251}}@{}}
\toprule\noalign{}
\begin{minipage}[b]{\linewidth}\raggedright
\textbf{Characteristic}
\end{minipage} & \begin{minipage}[b]{\linewidth}\centering
\textbf{Value}
\end{minipage} \\
\midrule\noalign{}
\endhead
\bottomrule\noalign{}
\endlastfoot
Region --- European Union & 23 (42.6\%) \\
Region --- North America & 19 (35.2\%) \\
Region --- Non-EU Europe (incl.~UK) & 7 (13.0\%) \\
Region --- Asia-Pacific / Global & 5 (9.3\%) \\
Years of professional experience --- median & 17 \\
Years --- range & 1--65 \\
Self-rated LLM understanding (1--7) --- mean & 4.89 \\
Self-rated deepfake familiarity (1--7) --- mean & 4.54 \\
Attribution preference --- anonymous & 27 (50.0\%) \\
Attribution preference --- acknowledgment only & 14 (25.9\%) \\
Attribution preference --- quote-attribution opt-in & 13 (24.1\%) \\
\end{longtable}
}

\begin{quote}
\emph{Note: Descriptive statistics, distributional diagnostics, recruitment details, and the domain-level policy priority matrix.}
\end{quote}

\emph{\textbf{Table C2.} Distribution shape of mitigation effectiveness ratings across strategies (N = 54).}

{\def\LTcaptype{none} 
\begin{longtable}[]{@{}
  >{\raggedright\arraybackslash}p{(\linewidth - 24\tabcolsep) * \real{0.2375}}
  >{\centering\arraybackslash}p{(\linewidth - 24\tabcolsep) * \real{0.0458}}
  >{\centering\arraybackslash}p{(\linewidth - 24\tabcolsep) * \real{0.0457}}
  >{\centering\arraybackslash}p{(\linewidth - 24\tabcolsep) * \real{0.0457}}
  >{\centering\arraybackslash}p{(\linewidth - 24\tabcolsep) * \real{0.0457}}
  >{\centering\arraybackslash}p{(\linewidth - 24\tabcolsep) * \real{0.0457}}
  >{\centering\arraybackslash}p{(\linewidth - 24\tabcolsep) * \real{0.0457}}
  >{\centering\arraybackslash}p{(\linewidth - 24\tabcolsep) * \real{0.0640}}
  >{\centering\arraybackslash}p{(\linewidth - 24\tabcolsep) * \real{0.0822}}
  >{\centering\arraybackslash}p{(\linewidth - 24\tabcolsep) * \real{0.0731}}
  >{\centering\arraybackslash}p{(\linewidth - 24\tabcolsep) * \real{0.1005}}
  >{\centering\arraybackslash}p{(\linewidth - 24\tabcolsep) * \real{0.0822}}
  >{\centering\arraybackslash}p{(\linewidth - 24\tabcolsep) * \real{0.0863}}@{}}
\toprule\noalign{}
\begin{minipage}[b]{\linewidth}\raggedright
\end{minipage} & \multicolumn{7}{>{\centering\arraybackslash}p{(\linewidth - 24\tabcolsep) * \real{0.3382} + 12\tabcolsep}}{%
\begin{minipage}[b]{\linewidth}\centering
\textbf{Rating}
\end{minipage}} & \begin{minipage}[b]{\linewidth}\centering
\end{minipage} & \begin{minipage}[b]{\linewidth}\centering
\end{minipage} & \begin{minipage}[b]{\linewidth}\centering
\end{minipage} & \begin{minipage}[b]{\linewidth}\centering
\end{minipage} & \begin{minipage}[b]{\linewidth}\centering
\end{minipage} \\
\begin{minipage}[b]{\linewidth}\raggedright
\textbf{Strategy}
\end{minipage} & \begin{minipage}[b]{\linewidth}\centering
\textbf{1}
\end{minipage} & \begin{minipage}[b]{\linewidth}\centering
\textbf{2}
\end{minipage} & \begin{minipage}[b]{\linewidth}\centering
\textbf{3}
\end{minipage} & \begin{minipage}[b]{\linewidth}\centering
\textbf{4}
\end{minipage} & \begin{minipage}[b]{\linewidth}\centering
\textbf{5}
\end{minipage} & \begin{minipage}[b]{\linewidth}\centering
\textbf{6}
\end{minipage} & \begin{minipage}[b]{\linewidth}\centering
\textbf{7}
\end{minipage} & \begin{minipage}[b]{\linewidth}\centering
\textbf{Mean}
\end{minipage} & \begin{minipage}[b]{\linewidth}\centering
\emph{\textbf{SD}}
\end{minipage} & \begin{minipage}[b]{\linewidth}\centering
\textbf{High \%}
\end{minipage} & \begin{minipage}[b]{\linewidth}\centering
\textbf{Mid \%}
\end{minipage} & \begin{minipage}[b]{\linewidth}\centering
\textbf{Low \%}
\end{minipage} \\
\midrule\noalign{}
\endhead
\bottomrule\noalign{}
\endlastfoot
Government regulation & 2 & 2 & 5 & 5 & 13 & 12 & 15 & 5.24 & 1.65 & 74.1 & 9.3 & 16.7 \\
Digital watermarking / provenance & 1 & 3 & 2 & 11 & 11 & 13 & 13 & 5.20 & 1.53 & 68.5 & 20.4 & 11.1 \\
Media / public literacy & 0 & 5 & 6 & 9 & 9 & 10 & 15 & 5.07 & 1.67 & 63.0 & 16.7 & 20.4 \\
Platform enforcement & 0 & 4 & 4 & 12 & 15 & 11 & 8 & 4.91 & 1.42 & 63.0 & 22.2 & 14.8 \\
Technical detection & 1 & 3 & 8 & 13 & 13 & 13 & 3 & 4.57 & 1.40 & 53.7 & 24.1 & 22.2 \\
\end{longtable}
}

\emph{Note: Likert votes (count per rating) per strategy, with summary share of high (≥ 5), mid (= 4), and low (≤ 3) ratings. All five distributions are right-skewed and unimodal. None exhibits the bimodal pattern that would indicate genuine effectiveness polarization. Disagreement in the forced-choice rankings therefore reflects priority-setting rather than effectiveness disagreement.}

\emph{\textbf{Table C3.} Recruitment funnel from candidate identification to valid responses.}

{\def\LTcaptype{none} 
\begin{longtable}[]{@{}
  >{\raggedright\arraybackslash}p{(\linewidth - 2\tabcolsep) * \real{0.5351}}
  >{\raggedright\arraybackslash}p{(\linewidth - 2\tabcolsep) * \real{0.0817}}@{}}
\toprule\noalign{}
\begin{minipage}[b]{\linewidth}\raggedright
\textbf{Stage}
\end{minipage} & \begin{minipage}[b]{\linewidth}\raggedright
\emph{\textbf{n}}
\end{minipage} \\
\midrule\noalign{}
\endhead
\bottomrule\noalign{}
\endlastfoot
Candidates identified & 516 \\
Directly contacted (any channel) & 454 \\
Channels --- Email & 214 \\
Channels --- LinkedIn & 95 \\
Channels --- Mastodon DM & 37 \\
Channels --- Bluesky / X DM & 23 \\
Channels --- Mixed / other & 85 \\
Survey responses received & 54 \\
Response rate (responses / contacted) & 11.9\% \\
\end{longtable}
}

\emph{Note: Snowball forwards from contacted candidates to colleagues are not captured in the denominator and are unknown}\emph{.}

\emph{\textbf{Table C4.} Policy priority matrix showing perceived threat by domain and modality (N = 54).}

{\def\LTcaptype{none} 
\begin{longtable}[]{@{}
  >{\raggedright\arraybackslash}p{(\linewidth - 6\tabcolsep) * \real{0.2500}}
  >{\raggedright\arraybackslash}p{(\linewidth - 6\tabcolsep) * \real{0.2500}}
  >{\raggedright\arraybackslash}p{(\linewidth - 6\tabcolsep) * \real{0.2500}}
  >{\raggedright\arraybackslash}p{(\linewidth - 6\tabcolsep) * \real{0.2500}}@{}}
\toprule\noalign{}
\begin{minipage}[b]{\linewidth}\raggedright
\textbf{Domain}
\end{minipage} & \begin{minipage}[b]{\linewidth}\centering
\textbf{Primary modality threat (\emph{M})}
\end{minipage} & \begin{minipage}[b]{\linewidth}\centering
\textbf{Priority lever}
\end{minipage} & \begin{minipage}[b]{\linewidth}\centering
\textbf{Supporting strategies}
\end{minipage} \\
\midrule\noalign{}
\endhead
\bottomrule\noalign{}
\endlastfoot
Political & Video deepfake (6.31) & Audiovisual provenance (e.g., C2PA) + platform enforcement & Statutory regulation of election-period synthetic media; rapid-response fact-checking; literacy targeted at electoral periods \\
Health & AI-generated text (5.80) & Labelling of AI-generated health content + medical fact-checking infrastructure & Clinician-facing literacy on LLM hallucinations; platform demotion of unverified medical claims; regulator coordination (e.g., FDA / EMA guidance) \\
Financial & Audio deepfake (5.65) & Voice-authentication standards for high-value transactions + KYC modernization & Sector-specific regulation of synthetic-voice fraud; cross-bank threat-intelligence sharing; staff training \\
Social (cross-cutting) & Mixed (video 6.00 / images 5.76) & Accessibility-first literacy reaching non-technical audiences & Provenance signals embedded in mainstream platforms; transparent platform moderation; community-driven counter-speech \\
\end{longtable}
}

\emph{Note: A preliminary, operational mapping from the descriptive threat patterns observed in this study to a prioritized mitigation stack per domain. The matrix is intended as a starting point for policymakers and platform operators rather than a closed framework; cells should be re-weighted as the AI capability landscape evolves. Cell content draws on respondent ratings (Findings) and on existing instruments \citep{bommasani2024fmti,chesney2019deepfakes,corsi2024synthetic,deangelis2023chatgpt,europarl2024aiact}. The matrix operationalizes the layered, contested-not-polarized mitigation landscape identified in the Implications section.}

\end{document}